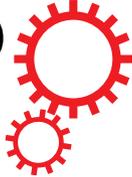

# A generalized architecture of quantum secure direct communication for *N* disjointed users with authentication

Ahmed Farouk[1,2], Magdy Zakaria[2], Adel Megahed[3] & Fatma A. Omara[4]

In this paper, we generalize a secured direct communication process between *N* users with partial and full cooperation of quantum server. So, $N-1$ disjointed users $u_1, u_2, ..., u_{N-1}$ can transmit a secret message of classical bits to a remote user $u_N$ by utilizing the property of dense coding and Pauli unitary transformations. The authentication process between the quantum server and the users are validated by *EPR* entangled pair and *CNOT* gate. Afterwards, the remained *EPR* will generate shared *GHZ* states which are used for directly transmitting the secret message. The partial cooperation process indicates that $N-1$ users can transmit a secret message directly to a remote user $u_N$ through a quantum channel. Furthermore, $N-1$ users and a remote user $u_N$ can communicate without an established quantum channel among them by a full cooperation process. The security analysis of authentication and communication processes against many types of attacks proved that the attacker cannot gain any information during intercepting either authentication or communication processes. Hence, the security of transmitted message among *N* users is ensured as the attacker introduces an error probability irrespective of the sequence of measurement.

The pioneering work of Bennett and Brassard[1] has been developed for the purpose of understanding quantum cryptography which is one of the most significant aspects of the laws of the quantum mechanics. This cryptography guarantees unconditional security[2–5] that is proven through the no-cloning theorem[6] as the transmitted quantum bit cannot be replicated or copied but its state can be teleported. Quantum teleportation and dense coding are the most often used quantum principles. In the former, the quantum information can be transmitted based on both classical communication and the maximally shared quantum entanglement among the distant parties[7–10]. In the latter, the classical information can be encoded and transmitted based on one quantum bit and on the maximally shared quantum entanglement among the distant parties as each quantum bit can transmit two classical bits[11,12]. A number of approaches and prototypes can be used to exploit quantum principles in order to secure communication between two parties and multiple parties[13–17]. While these approaches use different techniques to ensure private communication among authorized users, most of them still depend on the generation of secret random keys[18,19].

A quantum secret communication concept has recently been introduced; it is a kind of quantum communication in which secret messages can be transmitted through a quantum channel with or without additional classical communications[20–58]. A quantum secure direct communication (*QSDC*) transmits the secret messages directly between the communicating parties, from sender to receiver, without additional classical communication protocols, except for those used for the necessary eavesdropping check. In other words, the quantum key distribution (*QKD*) process and the classical communication of ciphertext are reduced into one single quantum communication procedure in *QSDC*. In *QSDC*, the concept of direct transmission of secret messages involves two kinds of meaning: On one hand, secret messages rather than raw keys are transmitted; on the other hand, the receiver does not require any separate classical communication from the sender to decode out secret messages[26,31,32,39–52,57,59]. In 2002, Long and Liu[52] put forward the first *QSDC* protocol, in which the secret message is transmitted directly. In the same year, by taking advantage of Einstein-Podolsky-Rosen (*EPR*) pairs as quantum information carriers, Boström and Felbinger[20] put forward the famous *QSDC* protocol referred to as the ping - pong protocol later. In[60], the authors enhanced the capability of the ping - pong protocol by adding two more unitary operations. In[22], a two-step quantum secure direct communication protocol was proposed for transferring quantum information by utilizing Einstein-Podolsky-Rosen (*EPR*) pair blocks to secure the transmission. In[21], the authentication and communication process was performed using Greenberg–Horne–Zeilinger (*GHZ*) states. First, the *GHZ* states were used for authentication purposes; the remaining *GHZ* states were used to transmit the secret message directly. In[17], the architecture for a centralized multicast scheme was proposed basing on a hybrid model of quantum key distribution and classical symmetric encoding. The proposed scheme solved the key generation and the management problem using a single entity called centralized Quantum Multicast Key Distribution Centre. In[61], a novel multiparty concurrent quantum secure direct communication protocol based on *GHZ* states and dense coding is introduced. In[59], a managed quantum secure direct communication protocol based on quantum encoding and incompletely entangled states is presented. In[57], a scheme for quantum secure direct dialogue protocols, which is adapted to both collective-dephasing noise and collective-rotation noise, is proposed by using the logical Bell states as the traveling states to resist collective noise. Different from *QSDC*, there is another kind of quantum secret communication named deterministic secure quantum communication (*DSQC*), where the receiver needs a separate classical communication from the sender to help decode out secret messages. In the framework of *DSQC*, the receiver can read out the secret message only after the transmission of at least one bit of additional classical information for each quantum bit, different from QSDC in which the secret message can be read out directly without exchanging any classical information[20,23,24,26,29,36,38,53,55,56,58]. In[53], a novel scheme for deterministic secure quantum communication (*DSQC*) over collective rotating noisy channel is proposed as a four special two-qubit states are found can constitute a noise-free subspaces, and so are utilized as quantum information carriers. In 2002, Beige et al.[54] first proposed a *DSQC* scheme based on single-photon two-quantum bit states. Then in 2004, Yan and Zhang[24] proposed a *DSQC* scheme based on *EPR* pairs and quantum teleportation. In 2005, Gao et al.[55] and Man et al.[26] proposed two *DSQC* protocols also based on entanglement swapping. In 2006, with EPR pairs based on the secret transmitting order of particles, Zhu et al.[20] proposed two *DSQC* schemes, one is a round trip scheme, and the other is a one way trip scheme. Lee et al.[29] proposed a protocol for controlled *DSQC* with Greenberger-Horne-Zeilinger (*GHZ*) states. In 2009, Xiu et al.[56] proposed a controlled *DSQC* scheme using five qubit entangled states and two-step security test. In[58], a hyper entangled Bell state is used to design faithful deterministic secure quantum communication and authentication protocol over collective-rotation and collective-dephasing noisy channel, which doubles the channel capacity compared with using an ordinary Bell state as a carrier; a logical hyper entangled Bell state immune to collective-rotation and collective-dephasing noise is constructed. Different quantum authentication approaches have been developed for preventing various types of attacks and especially man-in-the-middle attack[62–66].

[1]Information Technology Department, Al-Zahra College for Women, Oman. [2]Faculty of Computer and Information Sciences, Mansoura University, Egypt. [3]Faculty of Engineering, Cairo University, Egypt. [4]Faculty of Computers and Information Sciences, Cairo University, Egypt. Correspondence and requests for materials should be addressed to A.F. (email: dr.ahmedfarouk85@yahoo.com)

However, these quantum secure direct communication approaches are still prone to provide a low degree of effectiveness and an inadequate level of security. Here, we propose a convenient and efficient scheme for transmitting a series of classical messages among two, three, or more users (generalized to $N$ users). Therefore, $N − 1$ disjointed users $u_1, u_2, \ldots, u_{N−1}$ can transmit a secret message consisting of classical bits to a remote user $u_N$. The transmission process is accomplished by utilizing the property of dense coding and Pauli unitary transformations. First, the quantum server authenticates and verifies the identities of the communicated disjointed users through the generated entangled shared key and the *Controlled – NOT* gate. After the authentication is completed successfully, the remaining generated entangled shared key is used to generate shared *GHZ* states, which are intended for directly transmitting the secret message. If there is a quantum channel among the users, they can communicate using our partial cooperation process. In that case, $N − 1$ disjointed users generate a random sequence of bit strings of the transmitted plain message. Next, each user applies an appropriate unitary transformation according to his plain message bit string value and transmits the transformed message to $u_N$. Then, $u_N$ retrieves the original sent secret message by applying the $N – GHZ$ measurement to his/her particle and $u_1, u_2, \ldots, u_{N−1}$ particles. Afterwards, the quantum server calculates the status of his or her particle according to $x$ basis and announces his or her measurement results. Then, $u_N$ uses those measurements and the quantum server publication to retrieve the original sent secret bits by $u_1, u_2, \ldots, u_{N−1}$. If there is no quantum channel among the users, they can use our full cooperation process, but in this case the transformed message will be sent to the quantum server instead of $u_N$.

The efficiency and effectiveness of our protocol can be summarized into five points. First, the *GHZ* state is the maximally entangled state, so that the correlation can be more easily destroyed once any single $N$ particle is attacking. Second, using the $N – GHZ$ particle makes eavesdropping detection more effective and secure in comparison to some of the other *QSDC* protocols. For example[22], proposed a two-step quantum secure direct communication where an *EPR* pair block is used to transfer the secret message. Furthermore[34], proposed a multi-step quantum secure direct communication protocol where blocks of a multi-particle maximally entangled states are used to transmit secret messages. These protocols fail because the eavesdropper can capture some of the particles in the sequence and transmit what is left to the receiver through the quantum channel. If the eavesdropper intercepts the message sequence and conducts a *GHZ* measurement, he/she can retrieve some of the secret message. Therefore, the probability of information leakage exists. Third, our protocol increases the transmitted information capacity by using $N - GHZ$ states as these provide a large Hilbert space. Fourth, $N - 1$ users can transmit a particular message to the receiver, $u_N$, so the protocol is more effective as no quantum bits have to be discarded. Furthermore, the protocol is instantaneous as the receiver, $u_N$, is able to decode the message while receiving it and there is no additional classical exchange between $N - 1$ (sender) users and $u_N$ (receiver). Finally, the security analysis of the authentication and communication processes of our protocol against many types of attacks proves that our protocol is unconditionally secured and the attacker will not reveal any information about the key or the transmitted message in the case of directly calculating the transferred particles over the communicated channel from the quantum server to the disjoint user, and vice versa, as the attacker introduces an error probability irrespective of the sequence of measurement.

## Methods

**Bell States and *Controlled – NOT*.**   The Bell states are one of the main theories of quantum information processing that denote entanglement[67,68]. Bell states are specific, highly entangled quantum states of two particles denoted by *EPR*. There are many research groups that proposed different approaches for realizing and experimentally generating *EPR* states. In[69] a high-intensity source of polarization-entangled photon pairs can be realized with high momentum definition. The proposed scheme allowed ready preparation of all four of *EPR*-Bell states with two-photon fringe visibilities in excess of 97%. In[70] *EPR* can be experimentally setup by utilizing light pulse from a mode-locked Ti-sapphire laser through a frequency doubler. The ultraviolet pulse from the doubler is split into two beams by a balanced beam-splitter and is focused on four pairs of BBO crystals to provide four *EPR* photon pairs. In our scheme, *EPR* pair can be realized by utilizing the same concept introduced in[70].

These entangled particles have interrelated physical characteristics despite being spatially separated. When the quantum state is a multi-qubit, transformation can be achieved by applying the controlled quantum gates *CNOT* (*Controlled – NOT*), *FREDKIN* (*Controlled – SWAP*), and *TOFFOLI* (*Controlled – Controlled – NOT*). *CNOT* has an input of two qubits. It transforms the computational basis states by flipping the state of the second qubit only when the first qubit has a measurement of 1; otherwise, the quantum state remains unchanged[2,3,5,7,11]. The four Bell states (*EPR* pairs) used in both the authentication and communication processes in our scheme are defined by (Eq. 1). Authentication between the quantum server and users is achieved by the generated entangled shared key $|\Phi_{qu}^+\rangle$ and the *Controlled – NOT* gate. At the time of registration, the quantum server and disjoint user share a binary authentication key, $A_K$. Each sends one entangled particle to form an *EPR* pair, $|\Phi_{qu}^+\rangle$, in which the $q$ and $u$ particles correspond with the quantum server and disjoint user, respectively. The quantum server preserves $q$ at its location and transmits the $u$ particle to the intended disjoint user, as shown in (Eq. 2). Once the disjoint user obtains its $u$ particle, it prepares a new state particle, $n$ (See (Eq. 3), by encoding the shared authentication information according to the specified operation. When the quantum *CNOT* gate $\mathbb{C}_{OP}$ is performed on the transmitted particle and $n$, the resulted particle $r$ is a state of three entanglement particles (See Eq. 4). After applying the requested operation, the disjoint user keeps particle $u$ at its side and sends the resulted particle $|\Phi_r\rangle$ to the quantum server. Once the quantum server receives the resulted particle, $|\Phi_r\rangle$, it decodes it by applying a quantum *CNOT* gate $\mathbb{C}_{OP}$ on both the local particle $q$ and $n$ (See Eq. 6, 7). The quantum server verifies the identity of the disjoint user by measuring $|\phi_n\rangle$ on the basis of $Z$. The resulted state must measure at either 0 or 1. If the measurement is equal to $|A_{2i-1}, A_{2i}\rangle$, the disjoint user is authenticated. However, if the resulted measurement is erroneous—meaning that it is greater than the agreed threshold—then the authentication process will be terminated. Afterwards, the key is increased to authenticate the next disjoint user, sending the quantum server recursively back to step one until all disjoint users are authenticated.

$$|\Phi^\pm\rangle = \frac{1}{\sqrt{2}}\ (|00\rangle \pm |11\rangle), |\psi^\pm\rangle = \frac{1}{\sqrt{2}}\,(|01\rangle \pm |10\rangle) \tag{1}$$

$$|\Phi_{qu}^+\rangle = \frac{1}{\sqrt{2}}\,(|0_q 0_u\rangle + |1_q 1_u\rangle) \tag{2}$$

$$|\phi_n\rangle = |A_{2i-1} \otimes A_{2i}\rangle \tag{3}$$

| Value | First Bit | Second Bit | $u_i$ Transformation | GHZ Transformation |
|---|---|---|---|---|
| 0 | 0 | 0 | $I_{u_i}$ | $\frac{1}{\sqrt{2}}(\lvert\Phi^+\rangle_{ij}\lvert+\rangle_q + \lvert\Phi^-\rangle_{ij}\lvert-\rangle_q)$ |
| 1 | 0 | 1 | $X_{u_i}$ | $\frac{1}{\sqrt{2}}(\lvert\psi^+\rangle_{ij}\lvert+\rangle_q - \lvert\psi^-\rangle_{ij}\lvert-\rangle_q)$ |
| 2 | 1 | 0 | $Y_{u_i}$ | $\frac{1}{\sqrt{2}}(\lvert\psi^-\rangle_{ij}\lvert+\rangle_q - \lvert\psi^+\rangle_{ij}\lvert-\rangle_q)$ |
| 3 | 1 | 1 | $Z_{u_i}$ | $\frac{1}{\sqrt{2}}(\lvert\Phi^-\rangle_{ij}\lvert+\rangle_q + \lvert\Phi^+\rangle_{ij}\lvert-\rangle_q)$ |

**Table 1. Correlation between Received Classical Value and its Corresponding Unitary, *GHZ* Transformations and Quantum Bit Transformation Correlation during the Communication Process between Two Disjoint Users with Partial Cooperation of Quantum Server.**

| Value | First Bit | Second Bit | $u_i$ Transformation | GHZ Transformation |
|---|---|---|---|---|
| 0 | 0 | 0 | $I_{u_i}$ | $\frac{1}{\sqrt{2}}(\lvert\Phi^+\rangle_{iq}\lvert+\rangle_j + \lvert\Phi^-\rangle_{iq}\lvert-\rangle_j)$ |
| 1 | 0 | 1 | $X_{u_i}$ | $\frac{1}{\sqrt{2}}(\lvert\psi^-\rangle_{iq}\lvert+\rangle_j - \lvert\psi^-\rangle_{iq}\lvert-\rangle_j)$ |
| 2 | 1 | 0 | $Y_{u_i}$ | $\frac{1}{\sqrt{2}}(\lvert\psi^-\rangle_{iq}\lvert+\rangle_j - \lvert\psi^+\rangle_{iq}\lvert-\rangle_j)$ |
| 3 | 1 | 1 | $Z_{u_i}$ | $\frac{1}{\sqrt{2}}(\lvert\Phi^-\rangle_{iq}\lvert+\rangle_j + \lvert\Phi^+\rangle_{iq}\lvert-\rangle_j)$ |

**Table 2. Correlation between Received Classical Value and its Corresponding Unitary, *GHZ* Transformations and Quantum Bit Transformation Correlation during the Communication Process between Two Disjoint Users with Full Cooperation of Quantum Server.**

where $1 \leq i \leq N$ and $\otimes$ denotes the specified user operation.

$$\lvert\Phi_r\rangle = \mathbb{C}_{OP}(\lvert\phi_n\rangle \otimes \lvert\Phi_{qu}^+\rangle) \qquad (4)$$

where $\mathbb{C}_{OP} = \mathbb{C}_0$ at $A_{2i-1} = 0$ and $\mathbb{C}_{OP} = \mathbb{C}_1$ at $A_{2i-1} = 1$. $\mathbb{C}_0$ and $\mathbb{C}_1$ are described by (Eq. (5))

$$\mathbb{C}_0 = \lvert0\rangle\langle0\rvert \otimes I + \lvert1\rangle\langle1\rvert \otimes X,$$
$$\mathbb{C}_1 = \lvert+\rangle\langle+\rvert \otimes I + \lvert-\rangle\langle-\rvert \otimes X \qquad (5)$$

where $\lvert+\rangle = \frac{1}{\sqrt{2}}(\lvert0\rangle + \lvert1\rangle)$, $\lvert-\rangle = \frac{1}{\sqrt{2}}(\lvert0\rangle - \lvert1\rangle)$

$$\lvert\Phi_r'\rangle = \mathbb{C}_{OP}(\lvert\Phi_r\rangle) \qquad (6)$$

$$\lvert\Phi_r'\rangle = \mathbb{C}_{OP}\left(\lvert\phi_n\rangle \otimes \lvert\Phi_{qu}^+\rangle\right) \qquad (7)$$

**Quantum Bit Transformation.** Quantum computers can manipulate quantum information to transform a pure or mixed quantum state into another corresponding pure or mixed state[2,3,5,7,11]. In our scheme, the unitary transformation operations are defined by (Eq. (8)). For simplicity, we use $X$, $Y$, $Z$ instead of $\sigma_x$, $i\sigma_y$, $\sigma_z$, respectively. These are used to transform the *GHZ* state at the side of the sender(s) into an unreadable form that corresponds to the generated original classical message before transmitting it to the receiver.

$$\begin{aligned}
I &= \lvert0\rangle\langle0\rvert + \lvert1\rangle\langle1\rvert \\
X &= \lvert0\rangle\langle1\rvert + \lvert1\rangle\langle0\rvert \\
Y &= \lvert0\rangle\langle1\rvert - \lvert1\rangle\langle0\rvert \\
Z &= \lvert0\rangle\langle0\rvert - \lvert1\rangle\langle1\rvert
\end{aligned} \qquad (8)$$

Tables 1–4 describe the correlation between the received classical value and its corresponding unitary and *GHZ* transformations. Tables 1 and 2 illustrate the correlation that occurs during the communication process between two disjoint users ($u_i$, $u_j$) with partial and full cooperation of the quantum server, respectively. The $u_i$ generates a sequence of random bit strings of transmitted, plain message. According to each transmitted bits, (00, 01, 10, 11), the disjoint user, $u_i$, applies one of the unitary transformation operations, $\hat{U} = \{\hat{U}_1, \hat{U}_2, \hat{U}_3, \hat{U}_4\}$, which correspond to the four Pauli operations, $\{I, X, Y, Z\}$, respectively. Afterwards, the *GHZ* states convert according to the transmitted bits and the $u_i$ transformation

      

| Value | $u_i$ Bits | $u_j$ Bit | $u_i$ and $u_j$ Transformations | GHZ Transformation |
|---|---|---|---|---|
| 0 | 0 | 0 | 0 | $(I_{u_i} \otimes I_{u_j})$ | $\frac{1}{\sqrt{2}}(|\Psi^+\rangle_{ijl}|+\rangle_q + |\Psi^-\rangle_{ijl}|-\rangle_q)$ |
| 1 | 0 | 0 | 1 | $(I_{u_i} \otimes X_{u_j})$ | $\frac{1}{\sqrt{2}}(|\phi^+\rangle_{ijl}|+\rangle_q + |\phi^-\rangle_{ijl}|-\rangle_q)$ |
| 2 | 0 | 1 | 0 | $(X_{u_i} \otimes I_{u_j})$ | $\frac{1}{\sqrt{2}}(|\psi^+\rangle_{ijl}|+\rangle_q - |\psi^-\rangle_{ijl}|-\rangle_q)$ |
| 3 | 0 | 1 | 1 | $(X_{u_i} \otimes X_{u_j})$ | $\frac{1}{\sqrt{2}}(|\varphi^+\rangle_{ijl}|+\rangle_q - |\varphi^-\rangle_{ijl}|-\rangle_q)$ |
| 4 | 1 | 0 | 0 | $(Y_{u_i} \otimes I_{u_j})$ | $\frac{1}{\sqrt{2}}(|\psi^-\rangle_{ijl}|+\rangle_q - |\psi^+\rangle_{ijl}|-\rangle_q)$ |
| 5 | 1 | 0 | 1 | $(Y_{u_i} \otimes X_{u_j})$ | $\frac{1}{\sqrt{2}}(|\varphi^-\rangle_{ijl}|+\rangle_q - |\varphi^+\rangle_{ijl}|-\rangle_q)$ |
| 6 | 1 | 1 | 0 | $(Z_{u_i} \otimes I_{u_j})$ | $\frac{1}{\sqrt{2}}(|\Psi^-\rangle_{ijl}|+\rangle_q + |\Psi^+\rangle_{ijl}|-\rangle_q)$ |
| 7 | 1 | 1 | 1 | $(Z_{u_i} \otimes X_{u_j})$ | $\frac{1}{\sqrt{2}}(|\phi^-\rangle_{ijl}|+\rangle_q + |\phi^+\rangle_{ijl}|-\rangle_q)$ |

**Table 3. Correlation between Received Classical Value and its Corresponding Unitary, *GHZ* Transformations and Quantum Bit Transformation Correlation during the Communication Process among Three Disjoint Users with Partial Cooperation of Quantum Server.**

| Value | $u_i$ Bits | $u_j$ Bit | $u_i$ and $u_j$ Transformations | GHZ Transformation |
|---|---|---|---|---|
| 0 | 0 | 0 | 0 | $(I_{u_i} \otimes I_{u_j})$ | $\frac{1}{\sqrt{2}}(|\Psi^+\rangle_{ijq}|+\rangle_l + |\Psi^-\rangle_{ijq}|-\rangle_l)$ |
| 1 | 0 | 0 | 1 | $(I_{u_i} \otimes X_{u_j})$ | $\frac{1}{\sqrt{2}}(|\phi^+\rangle_{ijq}|+\rangle_l + |\phi^-\rangle_{ijq}|-\rangle_l)$ |
| 2 | 0 | 1 | 0 | $(X_{u_i} \otimes I_{u_j})$ | $\frac{1}{\sqrt{2}}(|\psi^+\rangle_{ijq}|+\rangle_l - |\psi^-\rangle_{ijq}|-\rangle_l)$ |
| 3 | 0 | 1 | 1 | $(X_{u_i} \otimes X_{u_j})$ | $\frac{1}{\sqrt{2}}(|\varphi^+\rangle_{ijq}|+\rangle_l - |\varphi^-\rangle_{ijq}|-\rangle_l)$ |
| 4 | 1 | 0 | 0 | $(Y_{u_i} \otimes I_{u_j})$ | $\frac{1}{\sqrt{2}}(|\psi^-\rangle_{ijq}|+\rangle_l - |\psi^+\rangle_{ijq}|-\rangle_l)$ |
| 5 | 1 | 0 | 1 | $(Y_{u_i} \otimes X_{u_j})$ | $\frac{1}{\sqrt{2}}(|\varphi^-\rangle_{ijq}|+\rangle_l - |\varphi^+\rangle_{ijq}|-\rangle_l)$ |
| 6 | 1 | 1 | 0 | $(Z_{u_i} \otimes I_{u_j})$ | $\frac{1}{\sqrt{2}}(|\psi^-\rangle_{ijq}|+\rangle_l + |\Psi^+\rangle_{ijq}|-\rangle_l)$ |
| 7 | 1 | 1 | 1 | $(Z_{u_i} \otimes X_{u_j})$ | $\frac{1}{\sqrt{2}}(|\phi^-\rangle_{ijq}|+\rangle_l + |\phi^+\rangle_{ijq}|-\rangle_l)$ |

**Table 4. Correlation between Received Classical Value and its Corresponding Unitary, *GHZ* Transformations and Quantum Bit Transformation Correlation during the Communication Process among Three Disjoint Users with Full Cooperation of Quantum Server.**

(See Supplementary Information 1) for the communication process between two disjointed users with partial and full cooperation of the quantum server).

Tables 3 and 4 describe the same correlation but among three disjointed users ($u_i$, $u_j$ and $u_l$) with partial and full cooperation of quantum server respectively. The difference that a new user $u_j$ generates a random sequences bits string of transmitted plain message. $u_j$ applies $I_{u_i}|\Psi\rangle$ or $X_{u_i}|\Psi\rangle$ according to the value of particle 0 or 1 respectively. Afterwards, the *GHZ* states will be converted according to transmitted bits, $u_i$ and $u_j$ transformations (see Supplementary Information 2 for Communication Process between Three Disjoint Users with Partial and Full Cooperation of Quantum Server).

### *GHZ* States, Measurement and Source.

A *GHZ* state is a certain type of maximally entangled quantum state that includes at least three qubits (particles). This kind of state was first examined by Greenberger, Horne, and Zeilinger in 1989[71]. The standard *GHZ* state is defined as *qubits* = 3; when *qubits* > 3, the *GHZ* state is defined by Eq. (9) seen below:

$$|GHZ\rangle = \frac{|0\rangle^{\otimes \text{qubits}} + |1\rangle^{\otimes \text{qubits}}}{\sqrt{2}} \tag{9}$$

In our scheme, when the quantum server receives a user(s) request for communication with another user, the quantum server distributes *GHZ* entanglement states among the involved participants' users in the communication process. Distribution is established after successful completion of the authentication process prior to the commencement of the communication process. The quantum server distributes all generated particles but holds one for itself. As a consequence, the quantum server and the participated users become entangled due to the presence of only one particle per distributed *GHZ* state. In addition, *GHZ* measurement is used by the receiver or quantum server, depending on the type of cooperation used during inter-user communication. Consistent with the *GHZ* measurement result, the receiver determines which unitary operations are used by the senders to transform the *GHZ* state according to the original generated classical message before it is transmitted to the receiver. By obtaining the applied unitary operations, the receiver can retrieve the original sent message. The Eight *GHZ* States are defined by (Eq. (10)).

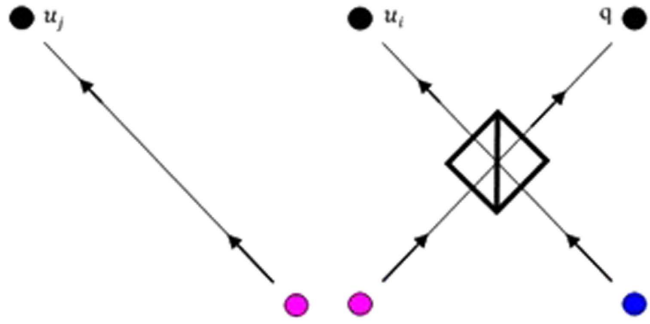

**Figure 1. Generation of *GHZ* States based on Einstein Podolsky-Rosen (*EPR*) "drawn by A.F".**

$$|\Psi^{\pm}\rangle = \frac{1}{\sqrt{2}}\left(|000\rangle \pm |111\rangle\right), \; |\psi^{\pm}\rangle = \frac{1}{\sqrt{2}}\left(|011\rangle \pm |100\rangle\right)$$

$$|\phi^{\pm}\rangle = \frac{1}{\sqrt{2}}\left(|010\rangle \pm |101\rangle\right), \; |\varphi^{\pm}\rangle = \frac{1}{\sqrt{2}}\left(|001\rangle \pm |110\rangle\right) \tag{10}$$

There're many research groups proposed different approaches for realizing and experimentally generating multi-photon *GHZ* states. In[72], an experimental entanglement of six-photon *GHZ* states, cluster states, and graph states is proposed. The generating of six-photon *GHZ* states and cluster states is achieved by *EPR*-entangled photon pairs. In[73] an effective protocol for preparation of $N$ – photon *GHZ* states with conventional photon detectors and can be realized through a simpler optical setup with a high success probabilities. In[74] a proposed a linear optical protocol to generate *GHZ* states of $N$ distant photons with certain success probabilities. The proposed set up involved simple linear optical elements $N$ pairs of the two-photon polarization entangled states, and the conventional photon detectors. In[75] an Experimental demonstration of five-photon entanglement and open-destination teleportation is proposed by utilizing two entangled photon pairs to generate a four-photon entangled state, which is then combined with a single-photon state. In[76] $N$ – particle *GHZ* states can be generated easily using the $N$ encoders preparation with cross-Kerr nonlinearities and can be realized simply through linear optical elements and homodyne detectors. In[70] Experimental generation of an eight-photon *GHZ* state is proposed. An eight-photon *GHZ* state with a measured fidelity of $0.59 \pm 0.02$ proved the presence of genuine eight-partite entanglement. This is achieved by improving the photon detection efficiency to 25% with a 300-mW laser pump. In our scheme, we use the same concept introduced in[70,72] to generate shared *GHZ* states. In order to develop the realization of $N$ – particle *GHZ* states, the photon detection efficiency has to be improved basing on laser pump. Furthermore, a large capacity of memory for all parties to store and retrieve the required information has to take into consideration.

After a successful completion of the authentication process between the quantum server and a specified user, the remaining *EPR* is used to generate shared *GHZ* states to transmit the secret message among communicated users directly. Figure 1 demonstrates how the *GHZ* states among $u_i$, the quantum server, and $u_j$ are generated according to the remaining *EPR*. Suppose, for example, the generated *EPR* for the authentication process between the quantum server and $u_i$ is given by (Eq. (11)). The quantum server particle is a part of another generated *EPR* for the purpose of authenticating $u_j$ (See Eq. 12). The result will then be a shared *GHZ* state among $u_i$, the quantum server, and $u_j$ (See Eq. 13). Similarly, $|\Psi^{+}_{iqjl}\rangle$, $|\Psi^{+}_{iqjlm}\rangle$, … and so on can be generated.

$$|\Phi^{+}_{iq}\rangle = \frac{1}{\sqrt{2}}\left(|0_i 0_q\rangle + |1_i 1_q\rangle\right) \tag{11}$$

$$|\Phi^{+}_{qj}\rangle\left(|0\rangle_i + |1\rangle_i\right) = \frac{1}{\sqrt{2}}\left(|0_q 0_j\rangle + |1_q 1_j\rangle\right)\left(|0\rangle_i + |1\rangle_i\right) \tag{12}$$

$$|\Psi^{+}_{iqj}\rangle = \frac{1}{\sqrt{2}}\left(|0_i 0_q 0_j\rangle + |1_i 1_q 1_j\rangle\right) \tag{13}$$

Tables 5–8 show the correlation between the quantum server publication, receiver measurement, sender operation(s), and sent bits. (Table 5) describes the correlation between the quantum server

| Quantum Server Publication | $u_j$ Measurement | $u_i$ Operation | Sent Bits |
|---|---|---|---|
| $|+\rangle_q$ | $|\Phi^+\rangle_{ij}$ | $I$ | 00 |
| $|+\rangle_q$ | $|\psi^+\rangle_{ij}$ | $X$ | 01 |
| $|+\rangle_q$ | $|\psi^-\rangle_{ij}$ | $Y$ | 10 |
| $|+\rangle_q$ | $|\Phi^-\rangle_{ij}$ | $Z$ | 11 |
| $|-\rangle_q$ | $|\Phi^-\rangle_{ij}$ | $I$ | 00 |
| $|-\rangle_q$ | $|\psi^-\rangle_{ij}$ | $X$ | 01 |
| $|-\rangle_q$ | $|\psi^+\rangle_{ij}$ | $Y$ | 10 |
| $|-\rangle_q$ | $|\Phi^+\rangle_{ij}$ | $Z$ | 11 |

**Table 5. Correlation between Quantum Server Publication, Receiver Measurement, Sender (s) Operation(s) and Sent Bits during partial cooperation process between Two Disjoint Users.** (C, D) shows an illustrative example for transmitting a message 100111 from $u_i$ to $u_j$ with partial and full support of quantum server respectively.

| Quantum Server Publication | $u_i$ Measurement | $u_j$ Operation | Sent Bits | $u_i$ Operation | Sent Bits | Message Sent |
|---|---|---|---|---|---|---|
| $|+\rangle_q$ | $|\Psi^+\rangle_{ijl}$ | $I$ | 00 | $I$ | 0 | 000 |
| $|+\rangle_q$ | $|\Phi^+\rangle_{ijl}$ | $I$ | 00 | $X$ | 1 | 001 |
| $|+\rangle_q$ | $|\psi^+\rangle_{ijl}$ | $X$ | 01 | $I$ | 0 | 010 |
| $|+\rangle_q$ | $|\varphi^+\rangle_{ijl}$ | $X$ | 01 | $X$ | 1 | 011 |
| $|+\rangle_q$ | $|\psi^-\rangle_{ijl}$ | $Y$ | 10 | $I$ | 0 | 100 |
| $|+\rangle_q$ | $|\varphi^-\rangle_{ijl}$ | $Y$ | 10 | $X$ | 1 | 101 |
| $|+\rangle_q$ | $|\Psi^-\rangle_{ijl}$ | $Z$ | 11 | $I$ | 0 | 110 |
| $|+\rangle_q$ | $|\Phi^-\rangle_{ijl}$ | $Z$ | 11 | $X$ | 1 | 111 |
| $|-\rangle_q$ | $|\Psi^-\rangle_{ijl}$ | $I$ | 00 | $I$ | 0 | 000 |
| $|-\rangle_q$ | $|\Phi^-\rangle_{ijl}$ | $I$ | 00 | $X$ | 1 | 001 |
| $|-\rangle_q$ | $|\psi^-\rangle_{ijl}$ | $X$ | 01 | $I$ | 0 | 010 |
| $|-\rangle_q$ | $|\varphi^-\rangle_{ijl}$ | $X$ | 01 | $X$ | 1 | 011 |
| $|-\rangle_q$ | $|\psi^+\rangle_{ijl}$ | $Y$ | 10 | $I$ | 0 | 100 |
| $|-\rangle_q$ | $|\varphi^+\rangle_{ijl}$ | $Y$ | 10 | $X$ | 1 | 101 |
| $|-\rangle_q$ | $|\Psi^+\rangle_{ijl}$ | $Z$ | 11 | $I$ | 0 | 110 |
| $|-\rangle_q$ | $|\Phi^+\rangle_{ijl}$ | $Z$ | 11 | $X$ | 1 | 111 |

**Table 6. Correlation between Quantum Server Publication, Receiver Measurement, Sender (s) Operation(s) and Sent Bits during partial cooperation process among Three Disjoint Users.**

publication, $u_j$ measurement, $u_i$ operation, and sent bits during a partial cooperation process between two disjoint users. The $u_j$ performs a Bell measurement on its particle and the $u_i$ particle; the quantum server then calculates the status of its particle according to the $x$ basis $\{+, -\}$ and announces the measurement results. The $u_j$ uses its measurements and the quantum server publications to retrieve the original secret bits sent by $u_i$. For example, when the $u_j$ measurement is equivalent to $|\psi^-\rangle$ and the quantum server publication is $|-\rangle$, the $u_j$ concludes that the $u_i$ applied an $X$ operation and the sent bits were 01 (See Supplementary Table S1) for the correlation during a full cooperation process between two disjoint users. Table 6 describes the same correlation of a partial process, but among three disjoint users. Here, $u_l$ performs a $GHZ$ measurement on its particle, $u_i$ particles, and $u_j$ particles. The quantum server calculates the status of its particle according to the $x$ basis $\{+, -\}$ and announces the measurement results. Then, $u_l$ uses its measurement and the quantum server publications to retrieve the original secret bits sent by both $u_i$ and $u_j$. For example, when the $u_l$ measurement is equivalent to $|\varphi^+\rangle$ and the quantum server publication is $|-\rangle$, $u_l$ can conclude that $u_i$ and $u_j$ applied $Y$ and $X$ operations, respectively, and the sent bits were 101 (See Supplementary Table S4) for correlation during a full cooperation process between three disjoint users). Tables 7 and 8 illustrate the transmission of messages 100111 from $u_i$ to $u_j$ with partial and full support of the quantum server, respectively.

| $u_i$ Plain Message | 10 | 01 | 11 |
|---|---|---|---|
| $u_i$ Operation | $Y$ | $X$ | $Z$ |
| GHZ Transformation | $\frac{1}{\sqrt{2}}(|\psi^-\rangle_{ij}|+\rangle_q - |\psi^+\rangle_{ij}|-\rangle_q)$ | $\frac{1}{\sqrt{2}}(|\psi^+\rangle_{ij}|+\rangle_q - |\psi^-\rangle_{ij}|-\rangle_q)$ | $\frac{1}{\sqrt{2}}(|\Phi^-\rangle_{ij}|+\rangle_q + |\Phi^+\rangle_{ij}|-\rangle_q)$ |
| $u_j$ Bell Measurement | $\psi^-$ | $\psi^+$ | $\Phi^+$ |
| Quantum Server Publication | $+$ | $+$ | $-$ |
| $u_j$ Retrieved Message | 10 | 01 | 11 |

**Table 7.  An illustrative example for transmitting a message 100111 from $u_i$ to $u_j$ with partial support of quantum server.**

| $u_i$ Plain Message | 10 | 01 | 11 |
|---|---|---|---|
| $u_i$ Operation | $Y$ | $X$ | $Z$ |
| GHZ Transformation | $\frac{1}{\sqrt{2}}(|\psi^-\rangle_{iq}|+\rangle_j - |\psi^+\rangle_{sq}|-\rangle_j)$ | $\frac{1}{\sqrt{2}}(|\psi^+\rangle_{iq}|+\rangle_j - |\psi^-\rangle_{iq}|-\rangle_j)$ | $\frac{1}{\sqrt{2}}(|\Phi^-\rangle_{iq}|+\rangle_j + |\Phi^+\rangle_{iq}|-\rangle_j)$ |
| Quantum Server Bell Measurement | $\psi^-$ | $\psi^+$ | $\Phi^+$ |
| $u_j$ Publication | $+$ | $+$ | $-$ |
| $u_j$ Retrieved Message | 10 | 01 | 11 |

**Table 8.  An illustrative example for transmitting a message 100111 from $u_i$ to $u_j$ with full support of quantum server.**

## Results and Discussion

**Masquerade as Dishonest Disjoint User Security Analysis.**   If an attacker would like to masquerade as dishonest disjoint user, then the attacker will work on the transmitting particle $u$ (*disjoint user particle*) from the quantum server to the disjoint user. With the assumption that the attacker applying a universal operation $\mathfrak{R}$ on $u$ see (Eq. (14, 15)).

$$|0_u \mathfrak{R}\rangle \rightarrow \alpha_0|0_u 0_a\rangle + \beta_0|0_u 1_a\rangle + \gamma_0|1_u 0_a\rangle + \delta_0|1_u 1_a\rangle \qquad (14)$$

$$|1_u \mathfrak{R}\rangle \rightarrow \alpha_1|0_u 0_a\rangle + \beta_1|0_u 1_a\rangle + \gamma_1|1_u 0_a\rangle + \delta_1|1_u 1_a\rangle \qquad (15)$$

where $|\mathfrak{R}\rangle$ represents an additional state which is created by the attacker, $a$ represents the attacker particle and,

$$|\alpha_0^2| + |\beta_0^2| + |\gamma_0^2| + |\delta_0^2| = |\alpha_1^2| + |\beta_1^2| + |\gamma_1^2| + |\delta_1^2| = 1 \qquad (16)$$

When the attacker applying its operation, a new shared key state will be created see (Eq. (17, 18)).

$$\left|\Phi_{qu}^+\right\rangle \rightarrow \left|\Phi_{qu}^{+\prime}\right\rangle \qquad (17)$$

$$|\Phi_{qu}^{+\prime}\rangle = \frac{1}{\sqrt{2}}(\alpha_0|0_q 0_u 0_a\rangle + \beta_0|0_q 0_u 1_a\rangle + \gamma_0|0_q 1_u 0_a\rangle + \delta_0|0_q 1_u 1_a\rangle$$
$$+ \alpha_1|1_q 0_u 0_a\rangle + \beta_1|1_q 0_u 1_a\rangle + \gamma_1|1_q 1_u 0_a\rangle + \delta_1|1_q 1_u 1_a\rangle)) \qquad (18)$$

The attacker transmits the new formulated state $|\Phi_{qu}^{+\prime}\rangle$ to the quantum server. Afterwards, the quantum server performs $\mathbb{C}_{OP}$ on the received state as a result one of four states $|\Phi_{qu}^{00}\rangle$, $|\psi_{qu}^{01}\rangle$, $|\psi_{qu}^{10}\rangle$, $|\Phi_{qu}^{11}\rangle$ which equivalent to the dual bits 00, 01, 10 and 11. With the assumption that the dual bits are equivalent to 00 then the quantum server operation is equivalent see (Eq. (19))

$$|\Phi_{qu}^{00}\rangle = \mathbb{C}_0|\Phi_{qu}^{+\prime}\rangle \qquad (19)$$

Through performing $\mathbb{C}_0$ the result is obtained by (Eq. (20))

$$|\Phi_{qu}^{00}\rangle = \frac{1}{\sqrt{2}} (\alpha_0 |0_q 0_u 0_a\rangle + \beta_0 |0_q 0_u 1_a\rangle + \gamma_0 |0_q 1_u 0_a\rangle + \delta_0 |0_q 1_u 1_a\rangle$$
$$+ \alpha_1 |1_q 0_u 1_a\rangle + \beta_1 |1_q 0_u 0_a\rangle + \gamma_1 |1_q 1_u 1_a\rangle + \delta_1 |1_q 1_u 0_a\rangle))$$

(20)

So the chance for discovering the attacker for $|\Phi_{qu}^{00}\rangle$ is $\acute{P}_{00}$ can be computed using (Eq. (20)) as by (Eq. (21))

$$\acute{P}_{00} = \frac{1}{2} (|\alpha_1{}^2| + |\gamma_1{}^2| + |\beta_0{}^2| + |\delta_0{}^2|)$$

(21)

As well when the dual bits are equivalent to 01, so the chance for discovering the attacker for $|\psi_{qu}^{01}\rangle$ is $\acute{P}_{01}$ can be computed using Eq. (18) as shown in (Eq. (22))

$$\acute{P}_{01} = \frac{1}{2} (|\alpha_0{}^2| + |\gamma_0{}^2| + |\beta_1{}^2| + |\delta_1{}^2|)$$

(22)

When the dual bits are equivalent to 10, so the chance for discovering the attacker of $|\psi_{qu}^{10}\rangle$ is $\acute{P}_{10}$ and equivalent to $\acute{P}_{01}$. Furthermore the chance for discovering the attacker of $|\Phi_{qu}^{11}\rangle$ is $\acute{P}_{11}$ and equivalent to $\acute{P}_{00}$. Accordingly, the total discovering probability of the attacker $\acute{P}_{Total}$ of each disjoint user is equivalent ½ see (Eq. (23, 24))

$$\acute{P}_{Total} = \frac{1}{4} (\acute{P}_{00} + \acute{P}_{01} + \acute{P}_{10} + \acute{P}_{11})$$

(23)

$$\acute{P}_{Total} = \frac{1}{2}$$

(24)

As stated by Simmons theory[77,78], the result of the previous equation proved that the proposed scheme is unconditionally secured under this type of attack.

**One-way Channel Substitution Fraudulent Attack Security Analysis.** As the transmitted particle from the quantum server to the disjoint user doesn't contain any fact about the authentication key, so in this type of attack only the restored $n$ (new state particle) from the disjoint user to the quantum server have to be measured. The maximum reachable information which an attacker may obtain over the communicated channel between a quantum server and a disjoint user can be computed by Holevo theory[79] see (Eq. (25))

$$\chi (\wp) = \varsigma (\wp) - \sum_i \mathcal{P}_i \varsigma (\wp_i)$$

(25)

As $\varsigma (\wp)$ is equivalent to Von Neumann entropy $-Tr (\wp \log_2 \wp)$, $\wp_i$ is a component in the hybrid status and $\mathcal{P}_i$ is the possibility of $\wp_i$ in the universe $\wp$. So the eavesdropper just has information about the authentication key by directly calculating the $n$ (new state particle), so the resulted $\chi (\wp)$ relies on the reduced density matrix of n, by substitution of (Eq. (25)) as shown in (Eq. (26))

$$\chi (\wp_n) = \varsigma (\wp_n) - \sum_i \mathcal{P}_i \varsigma (\wp_{ni})$$

(26)

As both $\wp_n$ and $\wp_{ni}$ required reduced density matrix for $\wp$ and $\wp_i$ respectively. For any authentication key, the reduced density matrix of n can be represented in (Eq. (27))

$$\wp_n = Tr_{qu}(|\Phi_r\rangle \langle \Phi_r|) = \frac{1}{2} I$$

(27)

In addition to, $\wp_{ni}$ is equivalent to the subsequent equations from ((28) to (32))

$$|\Phi_r^{00}\rangle = \frac{1}{\sqrt{2}} (|0_q 0_u 0_n\rangle + |1_q 1_u 1_n\rangle)$$

(28)

$$|\Phi_r^{01}\rangle = \frac{1}{\sqrt{2}} (|0_q 0_u 1_n\rangle + |1_q 1_u 0_n\rangle)$$

(29)

$$|\Phi_r^{10}\rangle = \frac{1}{\sqrt{2}}\left(|+_q +_u 1_n\rangle + |-_q -_u 0_n\rangle\right) \tag{30}$$

$$|\Phi_r^{11}\rangle = \frac{1}{\sqrt{2}}\left(|+_q +_u 0_n\rangle + |-_q -_u 1_n\rangle\right) \tag{31}$$

Therefore;

$$\mathbb{P}_{ni} = Tr_{qu}(|\Phi_r^i\rangle\langle\Phi_r^i|) = \frac{1}{2}I \tag{32}$$

By replacing values for both $\mathbb{P}_n$ and $\mathbb{P}_{ni}$ in Eq. (26), $\chi\ (\mathbb{P}_n) = 0$. So, the eavesdropper will not reveal any information about the key in case of directly calculating the transferred particles over the communicated channel from the quantum server to the disjoint user.

**Two- Way Channel Substitution Fraudulent Attack Security Analysis.** The attacker applies an operation $\Theta_1$ at his/her side on the transmitted particle $u$ and supportive particle $\mathcal{E}$. Afterwards, the attacker transmits the resulted particle to the disjoint user. When the disjoint user receives the transmitted particle, he/she does not realize that there is attacker and he did an operation. The disjoint user applyies his normal operation and transmits the resulted particle to the quantum server. The attacker intercepts the information particle sent by the disjoint user. The attacker applies an operation $\Theta_2$ at his side to the information particle and supportive particle $\eta$. Afterwards, the attacker transmits the resulted particle to the quantum server. The attacker attempts to retrieve certain amount of information about the key by employing two supportive particles $\varepsilon$ and $\eta$ (See Supplementary Information 3 for Full Calculations of Two - Way Channel Substitution Fraudulent Attack along with equations).

When the two-bit key $A_i\ A_{i+1} = 00$, so the resulting decoding state by the quantum server is given in (Eq. (33, 34))

$$|\Phi_{qu}^{00}\rangle = \mathbb{C}_0\Theta_2\{\mathbb{C}_0[\Theta_1(|\Phi_{qu}^+\rangle|\mathcal{E}\rangle)||\phi_n\rangle]|\eta\rangle\} \tag{33}$$

$$\begin{aligned}
|\Phi_{qu}^{00}\rangle = \frac{1}{\sqrt{2}}(&\alpha_\mathcal{E}\alpha_\eta|0_q0_u0_n\mathcal{E}_{00}\eta_{00}\rangle + \alpha_\mathcal{E}\beta_\eta|0_q0_u1_n\mathcal{E}_{00}\eta_{01}\rangle \\
+ &\beta_\mathcal{E}\beta_\eta|0_q1_u0_n\mathcal{E}_{01}\eta_{10}\rangle + \beta_\mathcal{E}\alpha_\eta|0_q1_u1_n\ \mathcal{E}_{01}\eta_{11}\rangle \\
+ &\beta_\mathcal{E}\alpha_\eta|1_q0_u1_n\mathcal{E}_{10}\ \eta_{00}\rangle + \beta_\mathcal{E}\beta_\eta|1_q0_u0_n\mathcal{E}_{01}\eta_{01}\rangle \\
+ &\alpha_\mathcal{E}\alpha_\eta|1_q1_u1_n\mathcal{E}_{11}\eta_{10}\rangle + \alpha_\mathcal{E}\alpha_\eta|1_q1_u0_n\mathcal{E}_{11}\eta_{11}\rangle
\end{aligned} \tag{34}$$

We can calculate the total possibility for discovering the attacker ($\mathbb{P}_{Total}$) in the authentication process is given by (Eq. (35))

$$\dot{\mathbb{P}}_{Total} = 1/2[\dot{\mathbb{P}}_{Total}(A_i = 0) + \dot{\mathbb{P}}_{Total}(A_i = 1)] \tag{35}$$

If the attacker would like to minimize his/her detection probability, he/she has to adjust $\mathbb{P}_{Total}$ as a minimum discovering probability (See Eq. 36) which is calculated under the condition of $\alpha_\mathcal{E} = \alpha_\eta = 1$

$$Total = Min(\dot{\mathbb{P}}_{Total}) = \frac{1}{4}(1 - \cos\ \theta_\mathcal{E}) \tag{36}$$

So, the attacker's total information amount on the transmitted key bits between the quantum server and the disjoint user can be estimated in (Eq. (37)).
where $\Theta_{Total}$ represents the total operation performed by the attacker $\Theta_1$ and $\Theta_2$, $x$ represents the key

$$\mathfrak{T}(A_K\ ,\Theta_{Total}) = \sum_{x,y}\mathbb{P}(A_K\ ,\Theta_{Total})\log_2\frac{\mathbb{P}(A_K\ ,\Theta_{Total})}{\mathbb{P}(A_K)\ \mathbb{P}(\Theta_{Total})} \tag{37}$$

values (00, 01, 10, 11) with probability $\mathbb{P}(x) = \frac{1}{4}$, $A_K$ indicates the selected random values from variable $x$, $y = \mathcal{E}_{ij}\eta_{\mu\tau}$ with $i, j, \mu,\ \tau \in \{0,\ 1\}$. Consequently, the joint gained information by attacker's total operation $\Theta_{Total}$ is given in (Eq. (38))

$$\mathfrak{T} = \frac{1}{4}[(1 + \sin\ \theta_\mathcal{E})\log_2(1 + \sin\ \theta_\mathcal{E}) + (1 + \sin\ \theta_\mathcal{E})\log_2(1 - \sin\ \theta_\mathcal{E})] \tag{38}$$

Since $\sin\ \theta_\mathcal{E} = \sqrt{8 \times Total - 16 \times Total^2}$ (see Supplementary Information (4) for Proving Relation between $\sin\ \theta_\mathcal{E}$ and $Total$), by substitution in Eq. (38)

$$\mathfrak{T} = \frac{1}{4}\Big[\Big(1 + \sqrt{8 \times Total - 16 \times Total^2}\Big)\log_2\Big(1 + \sqrt{8 \times Total - 16 \times Total^2}\Big)$$
$$+ \Big(1 - \sqrt{8 \times Total - 16 \times Total^2}\Big)\log_2\Big(1 - \sqrt{8 \times Total - 16 \times Total^2}\Big)\Big] \tag{39}$$

Therefore the total estimation probability $P_e$ of $A_K$ is given in (Eq. (40))

By simplification of Eq. (40) $P_e$ of $A_K$ is given in (Eq. (41)) (see Supplementary Information (5) for Proving Relation between $P_e$, $P$ and $\sin\theta_\varepsilon$)

$$P_e = \frac{(1 + \sin\theta_\varepsilon)}{2}\Big[\frac{1}{2}P + \frac{1}{4}(1 - P)\Big] + \frac{(1 - \sin\theta_\varepsilon)}{2}\Big[\frac{1}{4}(1 - P)\Big] \tag{40}$$

If $P = 1$ indicates that the total estimation probability $P_e$ is maximized see (Eq. (42)) (see Supplementary Information (6) for Proving Relation between $P_e$, $P_e{}^m$ and $Total$)

$$P_e = \frac{1}{8}\big[\,(\sin\theta_\varepsilon(3 \times P - 1) + 2]\big] \tag{41}$$

Therefore, the probability of the attacker for successfully retrieving the transmitted keys $P_e{}^r$ for $A_k = \{A_1, A_2, A_3 \dots\dots\dots\dots\dots\dots\dots\dots\dots\dots A_{2N}\}$ see (Eq. (43))

$$P_e{}^m = \frac{1}{4}\big(\sqrt{8 \times Total - 16 \times Total^2} + 1\big) \tag{42}$$

By substituting (Eq. (42)) in equation (Eq. (43)), so

Figure 2A is shown that the possibility for discovering the attacker while attempting to retrieve any information about the key bits is equal to non-zero. For example if $Total = 25\%$ means the attacker can

$$P_e{}^r = [P_e{}^m(1 - Total)]^{N/2} \tag{43}$$

gain maximal joint information $\mathfrak{T} = 0.5\ bit$ on the transmitted keys between the quantum server and

$$P_e{}^r = \Big[\frac{1}{4}\big(\sqrt{8 \times Total - 16 \times Total^2} + 1\big)(1 - Total)\Big]^{N/2} \tag{44}$$

disjoint user. From Fig. 2B, we can conclude that when the total estimation probability is maximized which means reaching to one, the attacker can positively maximum retrieve 0.5 bit of the transmitted key $A_k$ while the maximum total estimation probability $P_e{}^m$ is equal to 25%.

Figure 2C illustrates that if the minimum discovery probability is equal to [0, 12.5, 25, 50]% and $N = 2$, so the maximum values for successfully retrieving the information of the transmitted keys $A_k$ by the attacker are [0.25, $4.08 \times 10^{-1}$, $4.08 \times 10^{-1}$, $1.25 \times 10^{-1}$] respectively. Also, for $N = 16$ the maximum successfully for retrieving information of the transmitted keys $A_k$ by the attacker is [$1.53 \times 10^{-5}$, $7.7 \times 10^{-4}$, $3.91 \times 10^{-4}$, $5.96 \times 10^{-8}$] respectively. So while the number of the transmitted bits is increasing, the possibility for successful retrieval the transmitted information is decreased (See Supplementary Figure S6 (A–D) and Table S5 (A–D) for Full Calculations for Relation between $N$, $Total$, $P_e{}^r$).

Figure 2D shows that the maximum and minimum values for successfully retrieving the information of the transmitted keys $A_k$ by the attacker while $N = [2, 4, 8, 16]$ are [0.5, 0.25, 0.0625, $3.91 \times 10^{-3}$] and [0.0625, $3.91 \times 10^{-3}$, $1.53 \times 10^{-5}$, $2.32 \times 10^{-10}$] respectively. We can conclude that all maximum values are corresponding when $P_e{}^m$ and $Total$ is equal to [50, 0]%. So, the attacker can gain maximum information about the transmitted keys $A_k$ at this situation. Furthermore, the minimum value for the attacker to gain information about the transmitted keys $A_k$ is [$2.32 \times 10^{-10}$] corresponding when the value of $P_e{}^m$ and $Total$ is equal to [12.5, 50]%. (See Supplementary Figure S7 (A–D) and Table S6 (A–D) for Full Calculations for Relation between $P_e{}^r$ and $N$ while $P_e{}^m = [12.5, 25, 37.5, 50]\%$ and $Total = [0, 12.5, 25, 50\,]\%$).

Figure 2E demonstrates the maximum and minimum values for successfully retrieving the information of the transmitted keys $A_k$ by the attacker while $N = [2, 4, 8, 16]$ is [$1.87 \times 10^{-1}$, $3.51 \times 10^{-2}$, $1.23 \times 10^{-3}$, $1.53 \times 10^{-6}$] and [$1.56 \times 10^{-2}$, $2.5 \times 10^{-4}$, $5.96 \times 10^{-8}$, $3.55 \times 10^{-15}$] respectively. From Fig. 2E, we can conclude that all maximum and minimum values are corresponding when $P_e{}^m$ and $Total$ is equal to [50, 62.5]% and [12.5, 87.5]% respectively. So, the attacker can gain maximum information about the transmitted keys $A_k$ when $P_e{}^m$ and $Total$ equal to [50, 62.5]% and minimum when $P_e{}^m$ and $Total$ is equal to [12.5, 87.5]%. (See Supplementary Figure S8 (A–D) and Table S7 (A–D) for Full Calculations for Relation between $P_e{}^r$ and $N$ while $P_e{}^m = [12.5, 25, 37.5, 50]\%$ and $Total = [62.5, 75, 87.5]\%$).

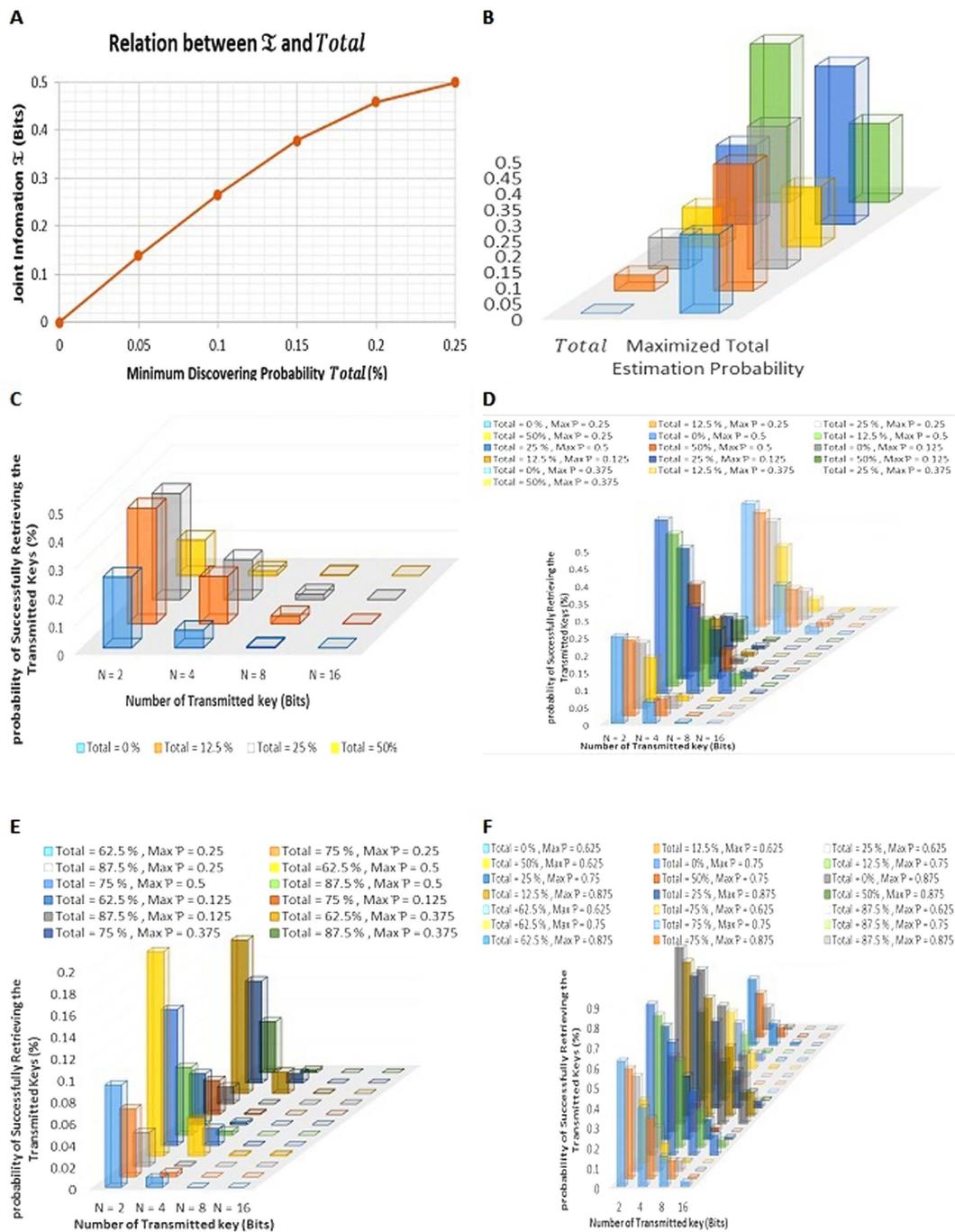

**Figure 2. Two- Way Channel Substitution Fraudulent Attack Security Analysis. (A)** The correlation between the joint information $\mathfrak{T}$ and the minimum discovering probability *Total*. **(B)** Correlation between Maximized Total Estimation Probability $P_e^{\,m}$ and the minimum discovering probability *Total*. **(C)** Relation between $P_e^{\,r}$, $N = [2, 4, 8, 16]$ and $Total = [0, 12.5, 25, 50]\%$. **(D)** Relation between $P_e^{\,r}$ and $N$ while $P_e^{\,m} = [12.5, 25, 37.5, 50]\%$ and $Total = [0, 12.5, 25, 50]\%$. **(E)** As Fig. 2D while $Total = [62.5, 75, 87.5]\%$. **(F)** Combined Fig. 2C,D while $P_e^{\,m} = [62.5, 75, 87.5]\%$. "drawn by A.F".

Figures 2F and 3 illustrate while the number of transmitted key bits $N$ becomes larger, the possibility of successfully retrieving $A_k$ becomes smaller and reach zero. So, the attacker will not reveal an enormous amount of information which can be ignored and avoided by updating key between the disjoint user and the quantum server periodically. In this case, the information of the attacker on the old key will be useless.

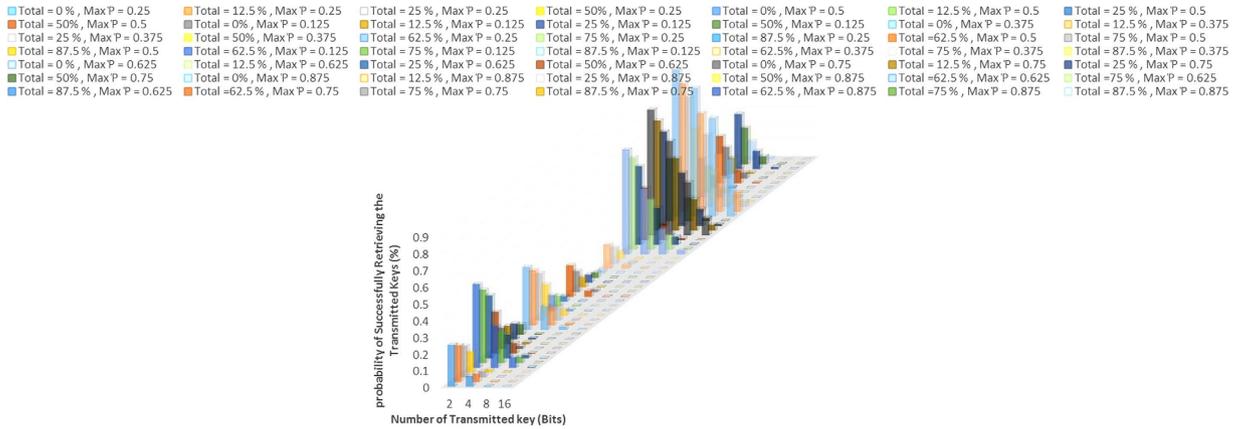

**Figure 3. Combined Fig. 2C–F "drawn by A.F".**

.....................................................................................................................................................

**Generalization of Communication Process between N Disjoint Users with Partial Cooperation of Quantum Server.** Here, we generalize our approach for Communication Process between $N$ Disjoint Users with Partial Cooperation of Quantum Server as shown in Fig. 4A. So, $N-1$ disjointed disjoint users $u_1$, $u_2$, ..., $u_{N-1}$ can transmit a secret message of classical bits to a remote user $u_N$ with partial cooperation of the quantum server. Firstly the quantum server distributes $N$ particles of $N+1$ particles $GHZ$ state $|GHZ\rangle_{1..........N} = \frac{1}{\sqrt{2}}(|\underbrace{00}_{N-1}...\underbrace{0\,0}_{q\ N}\rangle + |\underbrace{11}_{N-1}......\underbrace{1\,1}_{q\ N}\rangle)$ to $u_1$, $u_2$, ..., $u_N$. $u_1$, $u_2$, ..., $u_{N-1}$ chooses a randomly subset of $|GHZ\rangle_{1..........N}$ and keeps it confident. In addition, generate a random sequence bit strings of transmitted plain message. Next, each user applies appropriate unitary transformation according to his plain message. The bit string value $(\tilde{U}_1 \otimes \tilde{U}_2 ... \otimes \tilde{U}_{N-1}) \tilde{U}$ corresponds to four Pauli operations $\{I, X, Y, Z\}$. Afterwards, the selected $|GHZ\rangle_{1..........N}$ will be transformed according to $u_1$, $u_2$, ..., $u_{N-1}$ plain messages and their applied unitary transformations to $|GHZ\rangle_{1'..........N'} = \frac{1}{\sqrt{2}}(\underbrace{|GHZ\rangle_{N'}}_{\substack{u_1,...,u_N \\ Users}} \underbrace{|\pm\rangle_q}_{\substack{N+1 \\ quantum \\ server}} \pm \underbrace{|GHZ\rangle_{N''}}_{\substack{u_1,...,u_N \\ Users}} \underbrace{|\pm\rangle_q}_{\substack{N+1 \\ quantum \\ server}})$ where $\underbrace{|GHZ\rangle_{N'}}_{\substack{u_1,...,u_N \\ Users}}$ and $\underbrace{|GHZ\rangle_{N''}}_{\substack{u_1,...,u_N \\ Users}}$ are one of defined $GHZ$ states. Next, $|GHZ\rangle_{1'..........N'}$ transmitted to $u_N$, $u_N$ performs $N - GHZ$ measurement on his particle and $u_1$, $u_2$, ..., $u_{N-1}$ particles. Afterwards, the quantum server calculates the status of his particle according to $x$ basis $\{+, -\}$ and announces his measurement results. $u_N$ uses his measurements' and the quantum server publication for retrieving the original secret sent bits by $u_1$, $u_2$, ..., $u_{N-1}$. (See Supplementary Figures S1 and S3 for Communication Process between Two and Three Disjoint Users with Partial Support of Quantum Server respectively).

**Generalization of Communication Process between N Disjoint Users with Full Cooperation of Quantum Server.** Here, we generalize our approach for Communication Process between $N$ Disjoint Users with full Cooperation of Quantum Server as shown in Fig. 4B. The sequence of the steps is similar to a partial one except that the selected $|GHZ\rangle_{1..........N}$ will be transformed to $|GHZ\rangle_{1'..........N'} = \frac{1}{\sqrt{2}}(\underbrace{|GHZ\rangle_{N'}}_{\substack{u_1,...,q,\,u_{N-1} \\ Senders+ \\ quantum\,server}} \underbrace{|\pm\rangle_N}_{\substack{u_N \\ Reciever}} \pm \underbrace{|GHZ\rangle_{N''}}_{\substack{u_1,...,q,\,u_{N-1} \\ Senders+ \\ quantum\,server}} \underbrace{|\pm\rangle_N}_{\substack{u_N \\ Reciever}})$ where $\underbrace{|GHZ\rangle_{N'}}_{\substack{u_1,...,q,\,u_{N-1} \\ Senders+ \\ quantum\,server}}$ and $\underbrace{|GHZ\rangle_{N''}}_{\substack{u_1,...,q,\,u_{N-1} \\ Senders+ \\ quantum\,server}}$ are one of the defined $GHZ$ states. Next, $|GHZ\rangle_{1'..........N'}$ transmitted to the quantum server, quantum server performs $N - GHZ$ measurement on his particle and $u_1$, $u_2$, ..., $u_{N-1}$ particles. Afterwards, $u_N$ calculates the status of his particle according to $x$ basis $\{+, -\}$ and announces his measurement results. $u_N$ uses his publication and the quantum server measurement for retrieving the original sent secret bits by $u_1$, $u_2$, ..., $u_{N-1}$. (See Supplementary Figures S2 and S4 of Communication Process between Two and Three Disjoint Users with Full Support of Quantum Server respectively).

**Communication Process Security Analysis.** After $u_j$ retrieves the original secret sent bits which sent by $u_i$, $u_i$ informs $u_j$ about the positions of the transmitted particles and the selected unitary transformation applied to them. Afterwards, $u_j$ verifies the selected particles by $u_i$ and obtains an approximation of error percentage in the communication process. If the error percentage is under the specified threshold both $u_i$ and $u_j$ can continue the transmission of the secret messages, otherwise the communi-

**A**

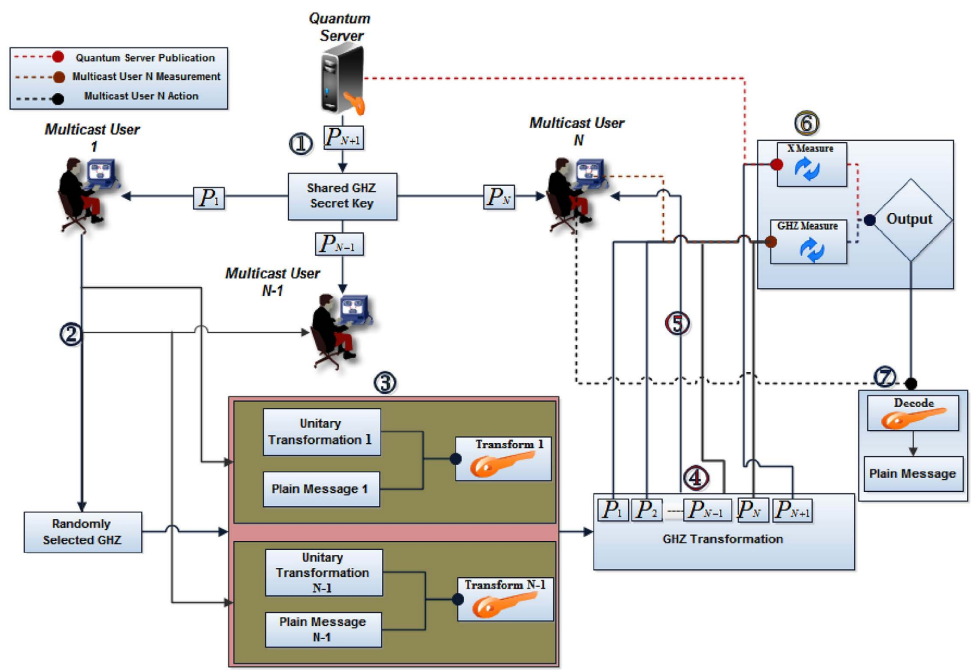

**B**

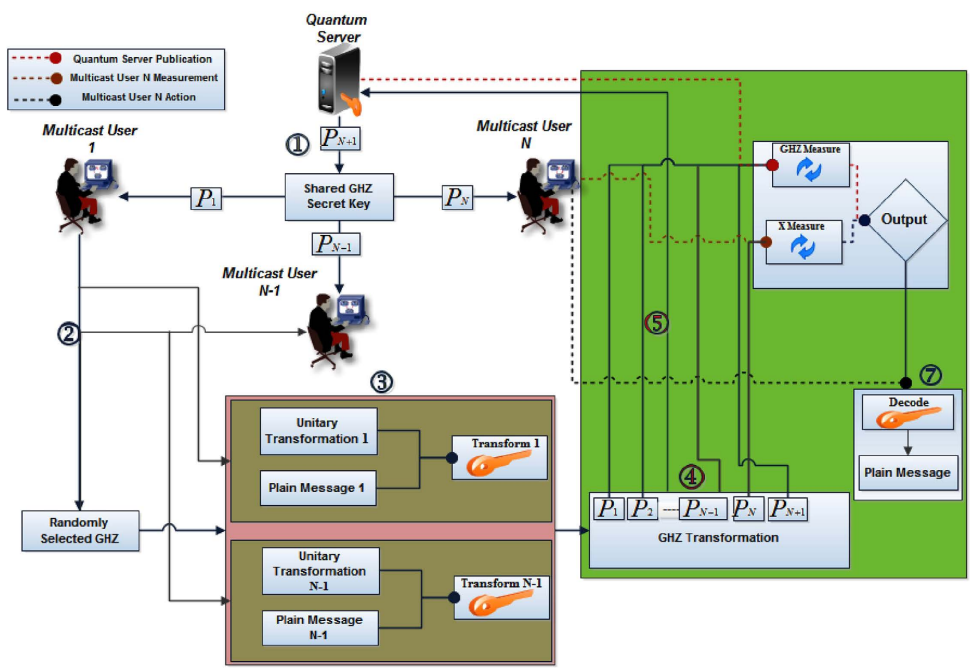

**Figure 4.** (**A**) Generalization of Communication Process between N Disjoint Users with Partial Cooperation of Quantum Server (**B**) with Full Cooperation of Quantum Server "drawn by A.F".

cation process will be terminated. If an attacker tries to spy on the transmitted *GHZ* particles, the attacker at most can obtain one particle. So, the attacker couldn't decide which operation is applied by $u_i$, consequently couldn't retrieve any transmitted secret bits. Suppose that the attacker apply an operation $\Theta_{u_iA}$ on $u_i$ and his qubit $|A\rangle$ see (Eq. (45)).

$$|0_A \mathcal{E}\rangle_{u_i A} \rightarrow \alpha_{\mathcal{E}} |0\rangle_{u_i} |\mathcal{E}_{00}\rangle_A + \beta_{\mathcal{E}} |1\rangle_{u_i} |\mathcal{E}_{01}\rangle_A$$

$$|1_A \mathcal{E}\rangle_{u_i A} \rightarrow \beta'_{\mathcal{E}} |0\rangle_{u_i} |\mathcal{E}_{10}\rangle_A + \alpha'_{\mathcal{E}} |1\rangle_{u_i} |\mathcal{E}_{11}\rangle_A \qquad (45)$$

Correspondingly, applying the unitary operation requires the following conditions $|\alpha_{\mathcal{E}}^2| + |\beta_{\mathcal{E}}^2| = 1$ and $|\alpha'^2_{\mathcal{E}}| + |\beta'^2_{\mathcal{E}}| = 1$.

So, the state of the protocol is transformed as follows, Firstly, the states after $u_i$ made a unitary operation see (Eq. (46)).

$$\begin{aligned} |\Psi_1\rangle_{iqjA} &= \Theta_{u_i} |\Psi\rangle_{iqj} \otimes |A\rangle_A \\ &= \tfrac{1}{2}(|000\rangle_{iqj} \mp |100\rangle_{iqj} + |011\rangle_{iqj} \pm |111\rangle_{iqj}) \otimes |A\rangle_A \end{aligned} \qquad (46)$$

Secondly, the states after the attacker applies a unitary transformation on $u_i$ and his qubit $|A\rangle$ see (Eq. (47, 48))

$$\begin{aligned} |\Psi_2\rangle_{iqjA} &= \Theta_{u_iA} |\Psi_1\rangle_{iqjA} \\ &= \tfrac{1}{2}\left[ \begin{pmatrix} |000\rangle_{iqj} (\alpha_{\mathcal{E}} |\mathcal{E}_{00}\rangle \pm \beta'_{\mathcal{E}} |\mathcal{E}_{01}\rangle)_A + |100\rangle_{iqj} (\beta_{\mathcal{E}} |\mathcal{E}_{01}\rangle \pm \alpha'_{\mathcal{E}} |\mathcal{E}_{11}\rangle)_A \\ + |011\rangle_{iqj} (\alpha_{\mathcal{E}} |\mathcal{E}_{00}\rangle \mp \beta'_{\mathcal{E}} |\mathcal{E}_{01}\rangle)_A + |111\rangle_{iqj} (\beta_{\mathcal{E}} |\mathcal{E}_{01}\rangle \mp \alpha'_{\mathcal{E}} |\mathcal{E}_{11}\rangle)_A \end{pmatrix} \right] \end{aligned} \qquad (47)$$

$$= \frac{1}{2\sqrt{2}} \left[ \begin{array}{l} |\phi_{ij}^+\rangle \left\{ \begin{array}{c} |+\rangle_q (\alpha_{\mathcal{E}} |\mathcal{E}_{00}\rangle \pm \beta'_{\mathcal{E}} |\mathcal{E}_{01}\rangle + \beta_{\mathcal{E}} |\mathcal{E}_{01}\rangle \mp \alpha'_{\mathcal{E}} |\mathcal{E}_{11}\rangle)_A \\ + \\ |-\rangle_q (\alpha_{\mathcal{E}} |\mathcal{E}_{00}\rangle \pm \beta'_{\mathcal{E}} |\mathcal{E}_{01}\rangle - \beta_{\mathcal{E}} |\mathcal{E}_{01}\rangle \pm \alpha'_{\mathcal{E}} |\mathcal{E}_{11}\rangle)_A \end{array} \right\} + \\[2.5em] |\phi_{ij}^-\rangle \left\{ \begin{array}{c} |+\rangle_q (\alpha_{\mathcal{E}} |\mathcal{E}_{00}\rangle \pm \beta'_{\mathcal{E}} |\mathcal{E}_{01}\rangle - \beta_{\mathcal{E}} |\mathcal{E}_{01}\rangle \pm \alpha'_{\mathcal{E}} |\mathcal{E}_{11}\rangle)_A \\ + \\ |-\rangle_q (\alpha_{\mathcal{E}} |\mathcal{E}_{00}\rangle \pm \beta'_{\mathcal{E}} |\mathcal{E}_{01}\rangle + \beta_{\mathcal{E}} |\mathcal{E}_{01}\rangle \mp \alpha'_{\mathcal{E}} |\mathcal{E}_{11}\rangle)_A \end{array} \right\} + \\[2.5em] |\Psi_{ij}^+\rangle \left\{ \begin{array}{c} |+\rangle_q (\alpha_{\mathcal{E}} |\mathcal{E}_{00}\rangle \mp \beta'_{\mathcal{E}} |\mathcal{E}_{01}\rangle + \beta_{\mathcal{E}} |\mathcal{E}_{01}\rangle \pm \alpha'_{\mathcal{E}} |\mathcal{E}_{11}\rangle)_A \\ - \\ |-\rangle_q (\alpha_{\mathcal{E}} |\mathcal{E}_{00}\rangle \mp \beta'_{\mathcal{E}} |\mathcal{E}_{01}\rangle - \beta_{\mathcal{E}} |\mathcal{E}_{01}\rangle \mp \alpha'_{\mathcal{E}} |\mathcal{E}_{11}\rangle)_A \end{array} \right\} + \\[2.5em] |\Psi_{ij}^-\rangle \left\{ \begin{array}{c} |+\rangle_q (\alpha_{\mathcal{E}} |\mathcal{E}_{00}\rangle \mp \beta'_{\mathcal{E}} |\mathcal{E}_{01}\rangle - \beta_{\mathcal{E}} |\mathcal{E}_{01}\rangle \pm \alpha'_{\mathcal{E}} |\mathcal{E}_{11}\rangle)_A \\ - \\ |-\rangle_q (\alpha_{\mathcal{E}} |\mathcal{E}_{00}\rangle \mp \beta'_{\mathcal{E}} |\mathcal{E}_{01}\rangle + \beta_{\mathcal{E}} |\mathcal{E}_{01}\rangle \pm \alpha'_{\mathcal{E}} |\mathcal{E}_{11}\rangle)_A \end{array} \right\} \end{array} \right] \qquad (48)$$

As shown in (Eq. (47, 48)), the attacker can't gain any information during intercepting the communication process. As well the attacker introduces an error probability of ½ irrespective of the sequence of measurement. For example, suppose that the attacker measurement is the same as $u_i$ means the applied unitary transformation may be $I$ or $Z$. If the measurements are different then the possible applied unitary transformation may be $X$ or $Y$.

## Acknowledgements

A.F. would like to thanks Josep Batle, Yan-Bing Li, Mosayeb Naseri, Nanrun Zhou and Ming-Ming Wang for their fruitful discussion and helpful comments.

## Author Contributions

A.F. and F.O. prepare authentication process, analysis of communication and authentication processes. A.F. prepares all figures and mathematics calculations. A.F., M.Z. and A.M. prepare the various types of communication process. All authors reviewed the manuscript and contributed for discussions of the project.

## Additional Information

**Supplementary information** accompanies this paper at http://www.nature.com/srep

**Competing financial interests:** The authors declare no competing financial interests.

**How to cite this article**: Farouk, A. *et al.* A generalized architecture of quantum secure direct communication for *N* disjointed users with authentication. *Sci. Rep.* **5**, 16080; doi: 10.1038/srep16080 (2015).